\documentclass[11pt,a4paper]{article}
\pdfoutput=1
\usepackage{jheppub}

\usepackage[utf8]{inputenc}
\usepackage[T1]{fontenc}
\usepackage{amsmath}
\usepackage{float}
\usepackage{graphicx}
\usepackage{comment}
\usepackage{xcolor}
\usepackage{color}
\usepackage{epsfig}
\usepackage{dcolumn}
\usepackage[hypcap=true]{caption}
\usepackage[hypcap=true]{subcaption}
\usepackage{bm}
\usepackage{amssymb}
\usepackage{slashed}
\usepackage{epstopdf}
\title{Higgs Flavor Violation as a Signal to Discriminate Models}

\author[a]{L.~de~Lima}
\author[a]{C.~S.~Machado}
\author[a]{R.~D.~Matheus}
\author[a]{L.~A.~F.~do~Prado}
\affiliation[a]{Instituto de F\'isica Te\'{o}rica, Universidade Estadual Paulista, SP, Brazil}

\emailAdd{leolima@ift.unesp.br, camilasm@ift.unesp.br, matheus@ift.unesp.br, leonidas@ift.unesp.br}

\abstract{
We consider the Higgs Lepton Flavor Violating process $h \rightarrow \tau \mu$, in which CMS found a $2.5~\sigma$ excess of events, from a model independent perspective, and find that it is difficult to generate this operator without also obtaining a sizeable Wilson coefficient for the dipole operators responsible for tau radiative decay, constrained by BABAR to $\mbox{BR}(\tau \rightarrow \mu \gamma)< 4.4 \times 10^{-8} $. We then survey a set of representative models for new physics, to determine which ones are capable of evading this problem.
We conclude that, should this measurement persist as a signal, type-III Two Higgs Doublet Models and Higgs portal-like models are favored, while SUSY and Composite Higgs models are unlikely to explain it.}


\begin{document}

\maketitle


\section{Introduction}

Now that we know of the existence of a Higgs boson with a mass of $125.7 \pm 0.4$ GeV \cite{Agashe:2014kda} the focus has shifted to the precise determination of its other properties, with the aim of determining if they are all consistent with the Standard Model (SM) Higgs boson. All the data collected so far is consistent with the predictions of the SM \cite{newCMS}, but the uncertainties are large and new physics could still be hidden, specially in the couplings of the Higgs boson. Deviations on these couplings could point to physics Beyond the SM (BSM), and since the Higgs couples directly to most of the other SM fields, it is an excellent probe for a multitude of different BSM models. New fermions not far above the electroweak scale, for instance, could alter the effective couplings of the Higgs boson with gluons and photons, and other sources of electroweak symmetry breaking could move its couplings to SM fermions away from the predicted values.

A lot of experimental effort has been put into measuring the decay rates both in the dominant and in subdominant channels (that are either forbidden at tree level or suppressed by small couplings). The subdominant decays are specially interesting, because the decay rates predicted by the SM are so small that they can be easily dominated by BSM physics, if any is present. In this paper, we will concentrate on Lepton Flavor Violating (LFV) decays of the Higgs: $h \rightarrow l_i l_j$ (which is understood as the sum of $h \rightarrow l_i^+ l_j^-$ and $h \rightarrow l_i^- l_j^+$). These decays are indirectly constrained by low energy data such as muon and tau rare decays, the $g-2$ and Eletric Dipole Moments (EDM) of electrons and muons.

The rare decays of the muon are so well tested that the ensuing constraints put the LFV decays of the Higgs to $e\mu$ beyond the reach of the LHC \cite{harnik}.
The constraints on tau decays, on the other side, are more allowing and recent papers suggested that FV decays of the Higgs into tau could be observed in the LHC \cite{harnik,diaz,ellis} with $\mbox{BR}(h\rightarrow\tau\mu) \lesssim 10^{-1}$. That possibility motivated a dedicated search for $h \rightarrow \tau \mu$ events, the results of which have been recently released by CMS \cite{cms}, with an excess of $2.5~\sigma$ \footnote{Which can be ambiguously read as either a fluctuation to set upper limits to the branching ratio, or as a signal and a measurement of the same branching ratio. We give more details on section \ref{subsecEFT}.}. The conclusions of \cite{harnik,diaz,ellis} were obtained in fairly model independent approaches, using effective Lagrangians and a small variety of operators. On the other side, many papers have explored Higgs FV in specific UV completions of the SM without any additional light particles \cite{adam,paper92,Arana-Catania:2013xma,susydiagramas,Arhrib:2012ax,Arganda:2014dta,Arroyo:2013kaa,sierra,andreas,heeck,lee,azatov,arganda05}
and some actually obtain, given the indirect constraints,  a much smaller $\mbox{BR}(h\rightarrow\tau\mu)$, below the current reach of the LHC \cite{adam,susydiagramas,susydiagramas,Arhrib:2012ax,Arganda:2014dta,Arroyo:2013kaa,azatov,arganda05}.
It is this apparent contradiction between model independent approaches and specific UV continuations that motivates this paper.

We will approach the problem, in section \ref{secMIC}, from a model independent perspective, using an effective theory and including dimension six and eight dipole operators, which have been mostly neglected in previous works on LFV Higgs decays. We will argue that the inclusion of those operators changes how the bounds from low energy data can be translated to constraints on LFV Higgs couplings, making the constraints in general more restrictive. We will also show that, in many realistic and simple UV continuations, dipole operators are generated with sizeable Wilson coefficients, and it is natural that in those cases the $h \rightarrow \tau \mu$ branching ratio will be much smaller than predicted by analyses that neglected those operators.

In section \ref{secModels} we will explore some representative classes of models to determine which ones generate which operators, and what are the relations between the Wilson coefficients. That classification will allow us to have a general view of the problem, separating models that predict branching ratios much smaller than the current experimental reach from those that are actually favored in case the $2.5~\sigma$ excess, observed by CMS in the $h \rightarrow \tau \mu$ channel, gets more significant with increasing luminosity.


\section{Model Independent Considerations}
\label{secMIC}

\newcommand{\efOp}[1]{\hat{\mathcal{O}}_{#1}}
\newcommand{\M}{\mathcal{M}}


\subsection{Effective Field Theory}
\label{subsecEFT}

In the Effective Field Theory (EFT) approach, we consider the most general Lagrangian compatible with the SM local symmetries and containing no
new degrees of freedom below a set scale $\Lambda$:
\begin{equation}
\mathcal{L}_{\text{EFT}}=\mathcal{L}^{(4)}_{\text{SM}}+\sum_{i>4} \mathcal{L}^{(i)},
\end{equation}
where $\mathcal{L}^{(4)}_{\text{SM}}$ is the renormalizable SM Lagrangian, and $\mathcal{L}^{(i)}$ contains the higher dimensional operators (of dimension $i$) generated by integrating out heavy fields above the scale $\Lambda$. A complete survey of the operators generated at dimension five and six can be found in \cite{Grzadkowski:2010es}.

We will be interested only in the leptonic sector, for which the SM Lagrangian is:
\begin{align}
\mathcal{L}^{(4)}_{\text{SM}}=\bar{L} i \slashed D L +\bar{E} i \slashed D E
- \left[ y_4 \bar{L} H E + \text{h.c}\right]+D_{\mu}H^{\dagger} D^{\mu}H-\lambda_H\left(H^{\dagger}H-\frac{v^2}{2} \right)^2,
\end{align}
where $L$ and $E$ are the lepton weak doublet and singlet fields (triplets in flavor space, indexes are suppressed), $H$ is the Higgs doublet, $D_\mu$ is the appropriate covariant derivative and $y_4$ is
a matrix in flavor space.

In the EFT, LFV Higgs interactions are mediated at lowest order by the dimension six operator\footnote{There are no 5-dimensional operators not involving the neutrino sector and other dimension six operators are either irrelevant to flavor violation or equivalent to the operator of eq. (\ref{3H}) \cite{harnik}.}
\begin{equation}
\mathcal{L}^{(6)} \supset \frac{c_{6 H}}{\Lambda^2}\efOp{6H} + \text{h.c} \equiv \frac{c_{6 H}}{\Lambda^2} \bar{L} H H^\dag H E + \text{h.c},
\label{3H}
\end{equation}
where $c_{6 H}$ is a matrix in flavor space and we define the operator $\efOp{6H} \equiv  \bar{L} H H^\dag H E$. In principle, the couplings for different chirality combinations may differ (e.g. $c_{6H}^{\tau_L \mu_R} \ne  c_{6H}^{\mu_L \tau_R}$) but, since we will be always looking at initial and final states containing both combinations, we consider a single coupling $c_{6 H}$, understood to be the average value $c_{6 H} \equiv \sqrt{c_{L R}^{2}+c_{R L}^{2}}$, where $LR$ and $RL$ are the chiralities of heavier and lighter fermions, in this order. This operator, in conjunction with the renormalizable Yukawa couplings ($y_4$) generates the effective flavor violating Yukawas, given in the mass basis by:
\begin{equation}
\sqrt{2}Y=y_4+ 3 {v^2 \over 2}  {c_{6 H} \over \Lambda^2}.
\end{equation}

In our analysis we assume that the diagonal entries of this matrix are close to their SM values (i.e., $Y_{i i}\simeq m_i/v$).

\begin{figure}[h]
\begin{subfigure}{.5\textwidth}
    \center
    \includegraphics[width=.5\linewidth]{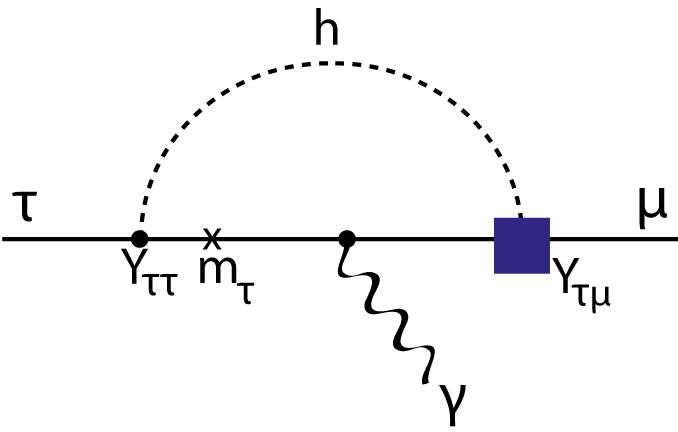}
    \caption{}
  \label{fig:c6Hloop}
\end{subfigure}%
\begin{subfigure}{.5\textwidth}
    \center
  \includegraphics[width=.5\linewidth]{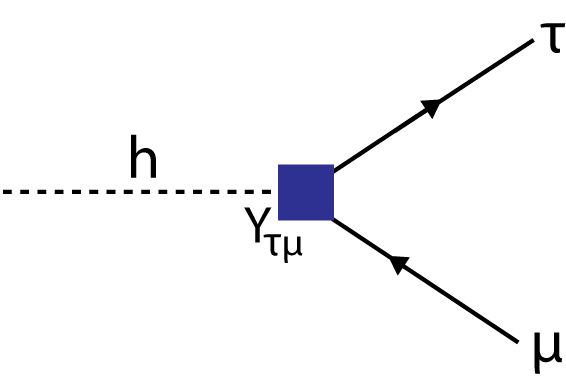}
   \caption{}
  \label{fig:taumu}
\end{subfigure}
\caption{(a) One-loop contribution to radiative lepton decay due to off-diagonal Yukawa couplings. (b) The LFV decay $h \rightarrow \tau \mu$.}
\label{f:diagrams}
\end{figure}

In the presence of a non-diagonal $Y$, a one-loop contribution to radiative lepton decays $l_i \rightarrow l_j \gamma$ is generated (see Fig. \ref{fig:c6Hloop}). If $\efOp{6H}$ is the only higher dimensional operator relevant for Higgs flavor physics, $Y_{i j}$ is constrained by the experimental bounds shown in Table \ref{tab:exp}. This diagram however must contain an insertion of the lepton Yukawa coupling, which greatly suppresses the radiative decay, allowing the LFV Higgs couplings $Y_{i j}$ to be sizeable while still respecting the limits in Table \ref{tab:exp}.

\begin{table}[h]
\begin{center}
\begin{tabular}{|r|l|}
\hline
~~Process~~ &~~Upper bound on BR~~\\
\hline \hline
~~$\mu \rightarrow \text{e} \gamma$ ~~& ~~$< 5.7 \times 10^{-13} $ ~~\cite{meg}\\
~~$\tau \rightarrow \text{e} \gamma$~~ & ~~$< 3.3 \times 10^{-8} $~~~ \cite{babar}\\
~~$\tau \rightarrow \mu \gamma$~~ & ~~$< 4.4 \times 10^{-8} $~~~ \cite{babar}\\
\hline
~~$h \rightarrow \tau \mu$~~ & ~~$< 1.57 \times 10^{-2}$ ~~\cite{cms}\\
\hline
\end{tabular}
\caption{Experimental bounds on LFV.}
\label{tab:exp}
\end{center}
\end{table}

Once a large enough $Y_{i j}$ is allowed, one may study the Higgs LFV decay $h \rightarrow l_i l_j$ generated by it (see Fig. \ref{fig:taumu}). In the case of $h \rightarrow \mu e$, the strong bound of $\mu\rightarrow e\gamma$ (Table \ref{tab:exp}) puts this process at a rate beyond the reach of the LHC: $\mbox{BR}(h\rightarrow \mu e) \lesssim 10^{-8}$ \cite{harnik}. The authors of \cite{harnik} claim that the non-diagonal Yukawas for $h \rightarrow \tau e$ and $h \rightarrow \tau \mu$ are still allowed to be big enough to produce a branching ratio at an observable level. Recently, CMS has released a data on the direct search for $h \rightarrow \tau \mu$  and there is an excess of events of a $2.5~\sigma$ significance in the channel \cite{cms}. That could be read as a statistical fluctuation and be used to put a bound of $\mbox{BR}(h\rightarrow\tau\mu) < 1.57\%$ at $95\%$ confidence level, or as a signal, which then leads to $\mbox{BR}(h\rightarrow\tau\mu) = \left( 0.89 \substack{+0.40 \\ -0.37}\right)\%$ \cite{cms}, only with more data will the situation become clear.
The possibility of confirming this as a signal is specially interesting because, as we will see in the following chapters, the conclusion of \cite{harnik} does not hold if there are other sources of flavor violation (besides $\efOp{6H}$), and a confirmation would disfavor models with such sources.
We will focus on this process for the remainder of this text.


\subsection{Interplay with Radiative LFV}
\label{subsecRLFV}

The conclusion of \cite{harnik}, that Higgs LFV decays may be observed in the tau sector, holds if the only contribution to the tau radiative decay comes from the Higgs one-loop diagram (Fig. \ref{fig:c6Hloop}). In general, however, other sources LFV above the scale $\Lambda$ may generate the dipole operators:
\begin{eqnarray}\label{dipole}
\mathcal{L}^{EFF}
& \supset &
e \frac{c_{6 \gamma}}{\Lambda^2} \efOp{6 \gamma} +e \frac{c_{8 \gamma}}{\Lambda^4} \efOp{8 \gamma} + \text{h.c},
\nonumber \\
& = &
e \frac{c_{6 \gamma}}{\Lambda^2} \bar{L} H \sigma_{\alpha\beta}E F^{\alpha\beta}+e \frac{c_{8 \gamma}}{\Lambda^4} \bar{L} H H^\dag H\sigma_{\alpha\beta}E F^{\alpha\beta}+ \text{h.c},
\end{eqnarray}
where $F^{\alpha\beta}$ is the electromagnetic field strength, $e$ is the electric charge and the dipole operators are defined to $\efOp{6 \gamma} \equiv \bar{L} H \sigma_{\alpha\beta}E F^{\alpha\beta}$ and $\efOp{8 \gamma} \equiv \bar{L} H H^\dag H\sigma_{\alpha\beta}E F^{\alpha\beta}$. Since these operators contribute at tree level to the flavor violating radiative decays they cannot be immediately dismissed. In the literature, one usually ignores $\efOp{8 \gamma}$
on account of it's larger suppression with the scale $\Lambda$ \cite{ellis,goudelis,davidson2,celis2}. However, in cases where $\efOp{6 \gamma}$
is not present (i.e., it is independently suppressed or unrelated to LFV, see the next section for examples), it may become the leading contribution to LFV radiative decays.

Thus, in general, the bounds from Table \ref{tab:exp} will apply to a combination of  ${c_{6 H} \over \Lambda^2}$, ${c_{6 \gamma} \over \Lambda^2}$ and ${c_{8 \gamma} \over \Lambda^4} $, generally making the restrictions on ${c_{6 H} \over \Lambda^2}$ stronger, unless we restrict the analysis to specific UV theories in which ${c_{6 \gamma} \over \Lambda^2}$ and ${c_{8 \gamma} \over \Lambda^4} $ are suppressed in relation to ${c_{6 H} \over \Lambda^2}$, either by having the dipole operators be generated a higher loop orders or by different physics at higher scales (making $\Lambda$ in eq.~(\ref{dipole}) different and bigger than $\Lambda$ in eq.~(\ref{3H})).

\begin{figure}[h]
\begin{subfigure}{\textwidth}
  \center
  \includegraphics[width=0.8\linewidth]{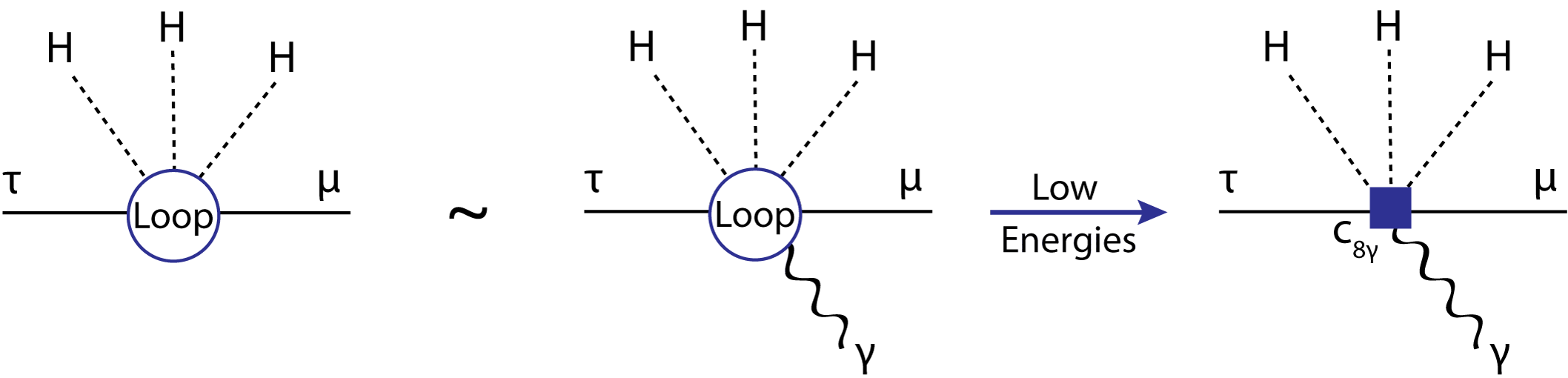}
   \caption{}
  \label{gen8}
\end{subfigure}\\
\begin{subfigure}{\textwidth}
    \center
    \includegraphics[width=0.8\linewidth]{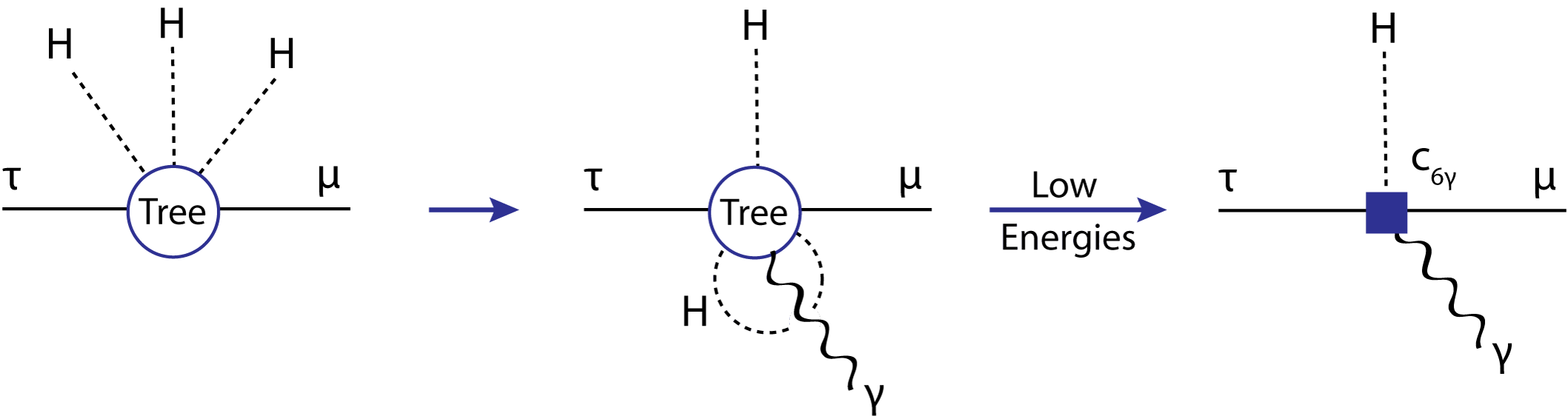}
    \caption{}
  \label{gen6}
\end{subfigure}
\caption{\label{gen68} Relation between the LFV Higgs coupling and higher dimensional dipole operators. (a) Obtaining the dimension eight dipole from
the loop generated LFV operator. (b) Obtaining the dimension six dipole from the tree generated LFV operator.}
\end{figure}

It turns out that it is quite hard to find simple UV completions that generate $c_{6 H} \gg c_{6 \gamma}$ and $c_{6 H} \gg c_{8 \gamma}$ at a given scale $\Lambda$.

For the sake of illustration, lets first assume that $\efOp{6H}$ is generated by the UV theory at loop level (one loop or more). If there is charge flowing through the loop, one can easily add an emitted photon, generating $\efOp{8 \gamma}$ with $c_{8 \gamma} \simeq c_{6 H}$ (see Fig. \ref{gen8}). The caveat here is that one could propose UV theories in which there is no charge going through the loop at leading order, in this case $c_{8 \gamma}$ would be suppressed by loop factors. We provide one such example in section \ref{subsecPortal}.
One could also obtain $\efOp{6 \gamma}$ from $\efOp{8 \gamma}$ by closing a loop with two external Higgs lines or by simply removing them. In the first case $c_{6 \gamma}$ will be a loop factor smaller than $c_{8 \gamma}$  and in the second case the relation between $c_{6 \gamma}$ and the other two couplings depends on the specific couplings of the Higgs with the particles going around the loop, making it effectively independent.

On the other side, if $\efOp{6 H}$ is present at tree level in the UV theory, one cannot simply attach a photon and get $\efOp{8 \gamma}$, because the diagram must vanish by Ward Identities. But now, one can obtain $\efOp{6 \gamma}$ by closing two of the Higgs legs in a loop and attaching a photon to that loop (see Fig. \ref{gen6}). In this case $c_{6 \gamma}$ will be smaller than $c_{6 H}$ by a loop factor, but since it contributes to the radiative decay at tree level while $c_{6 H}$ contributes only at loop level, the factors cancel and the dipole operator must be taken into account.

We see that one may connect the dipole operators to $\efOp{6H}$ and as such the Wilson coefficients $c_{6 \gamma},~c_{8 \gamma}$ are in general not independent from $c_{6 H}$. Of course, it is always possible to imagine that contributions from additional degrees of freedom to the dipole operators, as well as (approximate or exact) flavor symmetries, unrelated to $\efOp{6H}$  may greatly suppress the values of $c_{6 \gamma},~c_{8 \gamma}$ with respect to $c_{6 H}$, allowing the latter to be generated at lower scales and to have observable effects. In practice, however, a survey of the recent literature on the subject \cite{adam,paper92,Arana-Catania:2013xma,susydiagramas,Arhrib:2012ax,Arganda:2014dta,Arroyo:2013kaa,sierra,lee,azatov,arganda05} reveals that in many well motivated UV completions, such as SUSY and Composite Higgs scenarios, there is a great correlation between the sizes of $c_{6 \gamma},~c_{8 \gamma}$ and $c_{6 H}$, as we will discuss in the next section.

If the theory that UV completes the SM is a gauge theory, it is possible to show that the radiative dipole operators are always generated at one-loop level \cite{vodka}. This follows directly from imposing the QED Ward identities in the  radiative decay invariant amplitude. Such a consideration motivates extracting a universal $1/(16\pi^2)$ factor from the Wilson coefficients $c_{6 \gamma}$ and $c_{8 \gamma}$. Here we do not do this, because while these operators are always generated at loop level, the Wilson coefficient $c_{6 H}$ may be generated at tree or loop level, depending on the specific UV theory. As such, we will at first assume that $c_{6 H}$,  $c_{6 \gamma}$ and $c_{8 \gamma}$ are generated at the same order, implying that any loop factors are the same in both, to obtain our bounds. Later, when we discuss specific models, we will note the cases where $c_{6 H}$ is generated at tree level, or more generally at a lower loop order than the dipole operators, and hence is less suppressed by one or more loop factors.

Assuming that $c_{6 \gamma},~c_{8 \gamma}$ and $c_{6 H}$ are of the same order, we may compare the contributions to the $\tau \rightarrow \mu \gamma$ decay rate generated by each of the three higher dimensional operators. In the following expressions we neglect interference between channels, since our goal is to
find out in what regime only one of the Wilson coefficients is sizeable and the other two are small enough to be neglected. The contribution of $\efOp{6H}$ is shown in Fig. \ref{fig:c6Hloop} and the contributions of $\efOp{6\gamma}$ and $\efOp{8\gamma}$ are obtained by substituting $H$ by ${v \over \sqrt{2}}$ in the rightmost diagrams of Fig. \ref{gen68}.  The expressions are:
\begin{align}\label{c6G}
\Gamma_{6\gamma}&=\frac{ m_{\tau}^3}{16 \pi}\left(\frac{c_{6\gamma}}{\Lambda^2} e \frac{v}{\sqrt{2}}\right)^2,\\
\label{c6G2}
\Gamma_{8\gamma}&= \frac{ m_{\tau}^3}{16\pi}\left(\frac{c_{8\gamma}}{\Lambda^4}e\frac{v^3}{2\sqrt{2}}\right)^2,\\
\label{c6G3}
\Gamma_{6H}&=\frac{e^2}{4\pi}\frac{m_{\tau}^5}{64\pi^4}\left[\frac{c_{6H}  v^2}{\sqrt{2}\Lambda^2}\left(c^{\text{1-loop}}+c^{\text{2-loop}} \right)\right]^2,
\end{align}
where:
\begin{align}\label{eq:loops}
c^{\text{1-loop}} &\simeq \frac{1}{12 m_h^2}\frac{m_{\tau}}{v}\left[4-3 \log\left(\frac{m_h^2}{m_{\tau}^2} \right) \right],\\
\label{eq:loops2}
c^{\text{2-loop}} &\simeq  \frac{0.055}{m_h^2},
\end{align}
here we take the top Yukawa $y_{T}=(y_{T})_{\text{SM}}= m_T/v=0.70$.

We note that in the processes involving lepton radiative decay one must include a certain class of two loop diagrams, the so-called Barr-Zee diagrams~\cite{barr,harnik}. 
These diagrams, while of higher loop order, are less chirally suppressed, as such become parametrically more important than the diagram in Fig. \ref{fig:c6Hloop} and involve the same non-diagonal Yukawa coupling. The Barr-Zee contribution is denoted by $c^{\text{2-loop}}$ in eqs. (\ref{c6G3}) and (\ref{eq:loops2}). We also note that we are assuming that the one loop and two loop contributions have a positive relative sign i.e., there is no destructive interference between these diagrams. One way a cancellation may be achieved is if the top Yukawa (that enters in $c^{\text{2-loop}}$) has a negative sign with respect to the top mass, $y_T=-m_T \sqrt{2}/v$. In this case the numerical value of our bounds is changed, but qualitatively our results are unchanged.

\begin{figure}[h]
\center
\includegraphics[width=0.8\textwidth]{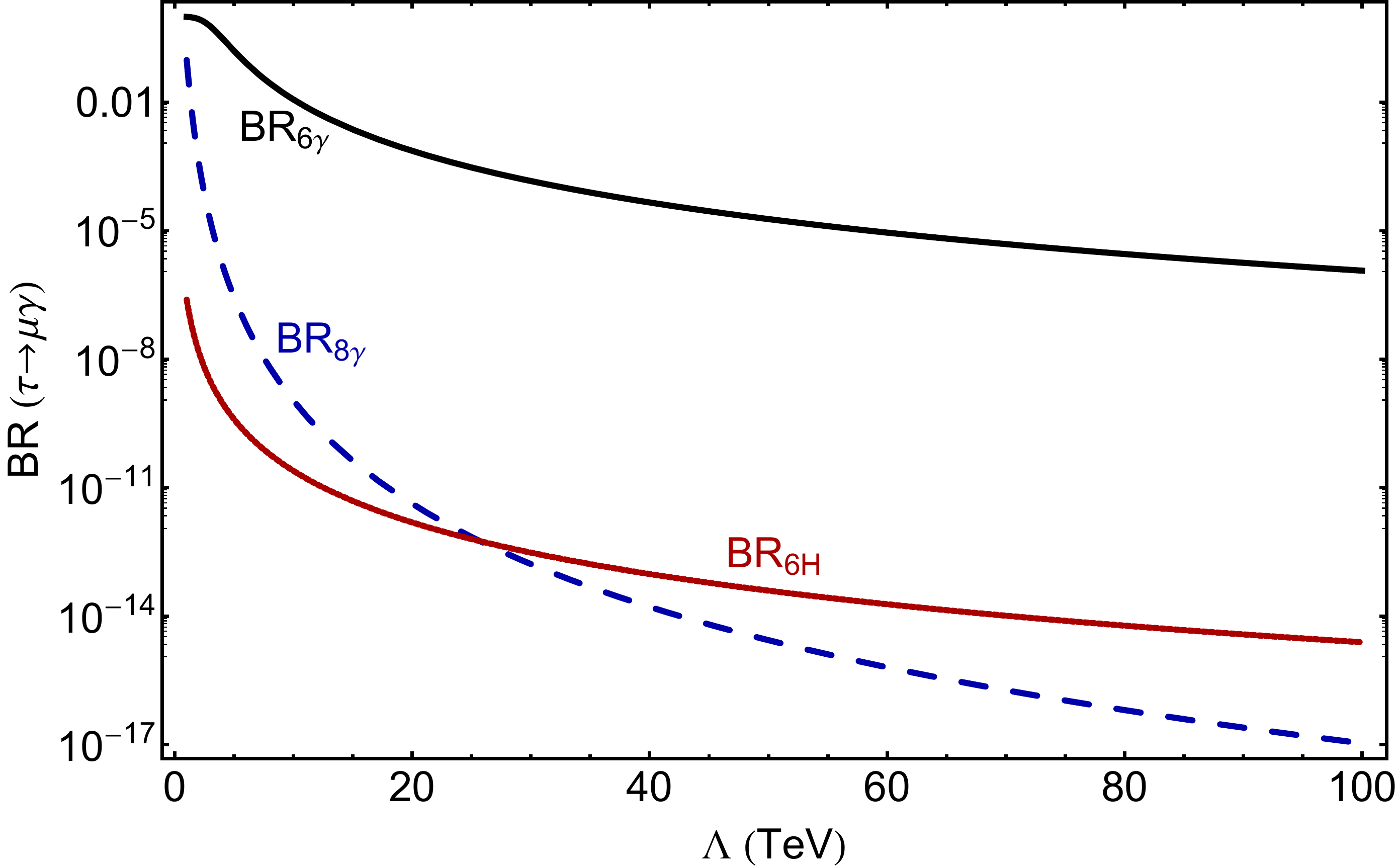}
\caption{\label{plotAmp} $\tau \rightarrow \mu \gamma$ branching ratio (including the Barr-Zee diagrams) from various operators as a function of the new physics scale $\Lambda$. The continuous black, dashed blue and dotted red lines are respectively the  branching ratios generated by  $\efOp{6\gamma}$, $\efOp{8 \gamma}$ and $\efOp{6H}$. Here, we consider $c_{6 \gamma}\sim c_{8 \gamma}\sim c_{6 H} \sim 1$, and interference effects are neglected.}
\end{figure}

In Fig. \ref{plotAmp}, we plot the branching ratio BR($\tau \rightarrow \mu \gamma$) calculated using eqs. (\ref{c6G}) to (\ref{c6G3}) as a function of the UV scale $\Lambda$. We see that, for similarly sized Wilson coefficients, $\efOp{6\gamma}$ dominates completely the amplitude, generating the strongest bound on $\Lambda$, while even the highly suppressed dimension eight operator $\efOp{8\gamma}$ becomes important for a cutoff below approximately 25 TeV and $c_{6 \gamma}\sim c_{8 \gamma}\sim c_{6 H} \sim 1$. In general $\efOp{8\gamma}$ will be less important than $\efOp{6H}$ for:
\begin{equation}
\Lambda \geq ( 25.7~\text{TeV})\sqrt{\frac{c_{8\gamma}}{c_{6H}}}.
\end{equation}

In the case where $\efOp{8\gamma}$ is generated at one higher loop order than $\efOp{6H}$, there is an extra factor of $\sim 1/(16\pi^2)$ in the definition of the operator, as discussed before. In this case the crossover happens at:
\begin{equation}
\Lambda \simeq (2.0~\text{TeV})\sqrt{\frac{c_{8\gamma}}{c_{6H}}}.
\end{equation}

It should not be so surprising that a dimension eight operator may play a role in such a case, since the diagram of Fig. \ref{fig:c6Hloop} that generates the dipole operator at one loop has an additional suppression due to the $\tau$ Yukawa insertion. In fact, it is for this same reason that we must include the Barr-Zee two loop diagrams when computing the contribution of the operator of eq. (\ref{3H}) to the tau decay rate. Hence, $\efOp{8\gamma}$ cannot be neglected in many cases of interest.

Once we have estimated the size of the various contributions to the LFV decay process, it becomes a relevant question to ask whether there are any interesting models in which the connection between the operators generating radiative and Higgs mediated LFV decay is broken, allowing the Higgs LFV decay to be seen at the LHC. One such example is a type-III two Higgs doublet model, explored in \cite{sierra}, while another one, explored recently in \cite{lee}, is provided by a scalar gauge singlet model, which may be a viable DM candidate. We discuss both these scenarios as well as the SUSY and Composite Higgs ones in the next section, keeping attention to the relative size of the higher dimentional LFV operators produced by each model
and whether or not they may generate the $h \rightarrow \tau \mu$ process at an observable level for the LHC.


\section{Results for Specific Models}
\label{secModels}


\subsection{Composite Higgs}
\label{subsecComposite}

In models where the EWSB happens due to the dynamics of a new strong sector, such as Randall Sundrum (RS)\cite{Randall:1999ee} models and their discrete versions (Quiver or $N$-site models \cite{michele,panico,quiver}), heavy vector-like leptons may mix linearly with the chiral SM  leptons, in the so-called partial compositeness scenario. In these models, the Higgs boson is a light composite bound state, and does not couple directly (i.e. in the flavor basis) to the SM leptons. Following a recent paper on a simplified model containing vector-like leptons \cite{adam}, one may write the Lagrangian
\begin{equation} \label{f:Lcomp}
\mathcal{L}=M \lambda_l \bar{L}_L \Psi_R +M \lambda_e \bar{E}_R \tilde{\Psi}_L -M c_l \bar{\Psi} \Psi-M c_e \bar{\tilde{\Psi}} \tilde\Psi+ Y \bar{\Psi}_L H \tilde{\Psi}_R+ \tilde{Y} \bar{\Psi}_R H \tilde{\Psi}_L +\mathrm{h.c.}
\end{equation}
where
$L_L$ and $E_R$ are the elementary electroweak lepton doublet and singlet, $\Psi$ and $\tilde{\Psi}$ are respectively the corresponding $SU(2)_L$ doublet and singlet vector-like fermions of the composite sector, $H$ is the Higgs doublet, $M$ is the mass scale of the heavy resonances, $\lambda_l,~\lambda_e,~Y, ~\tilde{Y}$ are matrices in flavor space (all fermion fields are flavor triplets) and we are in the $\{\Psi,\tilde{\Psi}\}$ basis where $c_l$ and $c_e$ are flavor diagonal.

\begin{figure}[t]
\center
\includegraphics[width=0.70\textwidth]{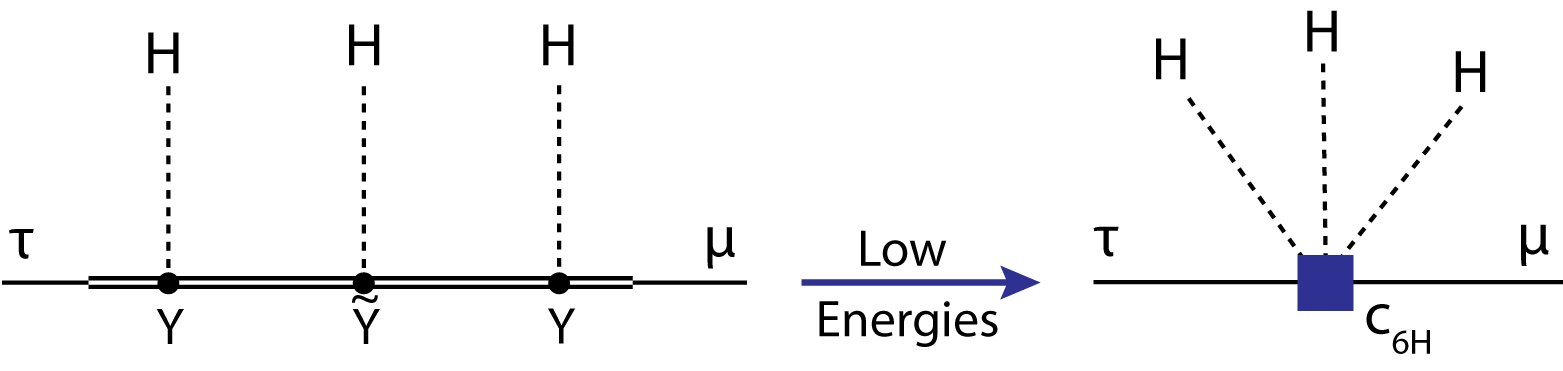}
\caption{Tree-level insertions of the Higgs generating $\efOp{6H}$ in the composite sector model of \cite{adam}. The double line represents vector-like heavy lepton states, while the thin lines are chiral elementary fermions.}
\label{diagramaadam}
\end{figure}

In this class of models, $\efOp{6H}$ is generated at tree level by integrating out the heavy fermions, as per the diagram of Fig. \ref{diagramaadam}.
It is essential in obtaining this operator that there be non-zero ``wrong'' chirality couplings $\tilde{Y}$ of the Higgs to the composite sector. Such an operator is naturally obtained when considering the brane localized (pure composite) Higgs as a limit of a bulk localized Higgs \cite{azatov}. Upon closing two of the external Higgs legs and adding an external photon, one generates $\efOp{6\gamma}$  and it is thus easy to see that the Wilson coefficients $c_{6\gamma}$ and $c_{6H}$ are proportional (as in Fig. \ref{gen68}). Furthermore, because the required chirality flip occurs on a vector-like fermion line, it does not suffer from chiral suppression by the tau Yukawa. Because of this strong correlation, the stringent bounds on the radiative decay $\tau \rightarrow \mu \gamma$ force the Higgs decay $h \rightarrow \tau \mu$ to be well below observational level. The authors of \cite{adam} obtain $\mbox{BR}(h\rightarrow\tau\mu) < 8.6 \times 10^{-6}$, which is much more restrictive than the current bounds obtained by direct searches: $\mbox{BR}(h\rightarrow\tau\mu) < 1.57 \times 10^{-2}$ \cite{cms}.

In a more complete model, such as the RS model explored in \cite{tanedo}, there are additional contributions to $c_{6\gamma}$ and $c_{8\gamma}$, due to composite gauge bosons (electroweak Kaluza-Klein (KK) modes). In this case, the main contribution is due to the first charged resonance $W^\prime$, which sets the strongest bound on the KK scale, of $\Lambda \simeq 5~$TeV for $Y,~ \tilde{Y}$ of order one. Barring cancellations with the typically smaller contributions from the neutral sector ($Z^\prime$ and $h$), one may use such a bound on the KK scale to obtain the expected value of the $h \rightarrow \tau \mu$ branching ratio to be of order $m_h/(8 \pi \Lambda^4) \sim 10^{-14}$, for $c_{6H} \simeq 1$.

Both these cases consider the Higgs to be a bound state of the strong sector that is accidentally light. A well motivated proposal to explain the lightness of the Higgs with respect to the other resonances is that it is a pseudo Nambu-Goldstone Boson (pNGB). In this case, one assumes the strong sector has a large global symmetry $\mathcal{G}$, spontaneously broken down to $\mathcal{H}$, such that the NGBs parametrizing the $\mathcal{G}/\mathcal{H}$ coset contain the Higgs doublet, conveniently represented by the non-linear sigma model field $\Sigma=\exp{(i h/f)}$, where $h$ are the components of the doublet $H$ and $f$ is the Higgs decay constant. The symmetry $\mathcal{G}$ is then broken by the Yukawa couplings and by gauging a subgroup $\mathcal{H}^\prime \supset SU(2)_L \times U(1)_Y$, giving the Higgs, at the one loop level, a small mass compared to the strong sector resonances.

The requirement that the Higgs couplings must respect a global symmetry of the strong sector, in which the SM gauge symmetry is embedded, modifies the effective Lagrangian, such that $\efOp{6H}$
must be aligned to the dimension four Yukawa coupling, suppressing flavor violation \cite{contino} \footnote{ While \cite{contino} considers the quark sector, the lepton sector is analogous. See \cite{micheleL,quiverL}.}.

Following \cite{contino}, this alignment can be seen by promoting the strong sector operators $\Psi,~\tilde{\Psi}$ to full representations of $\mathcal{G}$. Then, the Yukawa couplings $\lambda_l,~\lambda_e$ of eq. (\ref{f:Lcomp}) are made formally invariant under $\mathcal{G}$ by embedding the elementary fields in incomplete $\mathcal{G}$ representations, where the additional degrees of freedom are spurions. In place of the direct couplings to $H$, the strong sector must couple $\mathcal{G}$ invariantly to $\Sigma$.

Global invariance then forces the couplings to the doublet $H$ to take the schematic form
\begin{equation}
f \bar{L}_L \frac{H}{f} \left( y_4 + y_6  \frac{H^\dagger H}{f^2} +\cdots  \right) E_R \equiv \bar{L}_L P(\Sigma)E_R,
\end{equation}
where $P(\Sigma)$ is a polynomial in $\Sigma$, projected over the elementary fields. As long as the elementary fields are coupled to a single representation (e.g. the \textbf{5} of $SO(5)$ in the MCH5  model \cite{minimalcomposite}, in which case $P(\Sigma)= \Sigma \Sigma^T \rightarrow \sin(h/f)\cos(h/f)$), then all the higher dimensional non-derivative operators in $H$ are aligned and can be simultaneously diagonalized.

In this case the leading contribution comes from the kinetic mixing generated by integrating out the heavy leptons, which is typically smaller than (\ref{3H}) and leads to even lower Higgs LFV decay rates. One may then conclude that the observation of the process $h \rightarrow \tau \mu$ at a rate near the current experimental sensitivity would disfavour this class of models.


\subsection{2HDM}
\label{subsec2HDM}

In models with two electroweak Higgs doublets, the Yukawa Lagrangian for the charged leptons is written as
\begin{equation}
\mathcal{L}=y_1 \bar{L} \Phi_1 E + y_2 \bar{L} \Phi_2 E +\mathrm{h.c},
\end{equation}
where $\Phi_1$ and $\Phi_2$ are the Higgs doublets and $y_1,~y_2$ their corresponding Yukawa matrices.
Neglecting CP violation for simplicity, the neutral CP even scalar mass eigenstates are related to those in the flavor basis by
\begin{align}\label{f:2HDMrot}
h&=\sin(\alpha-\beta)\phi_1+\cos(\alpha-\beta)\phi_2 \nonumber, \\
H&=\cos(\alpha-\beta)\phi_1-\sin(\alpha-\beta)\phi_2.
\end{align}

We consider the case where the lightest eigenstate $h$ is the $125~$GeV Higgs. Limits on deviations from the SM couplings then demand $\sin(\alpha-\beta)\simeq 1$, while flavor constraints on B meson physics demand $m_H \geq 300~$GeV \cite{2hdmlimits}. If there is a parity symmetry distinguishing the two doublets, (e.g. type II 2HDM), one may forbid the couplings of $\Phi_2$ to the charged leptons at tree level. This case is closely related to the SUSY models considered in the following section and will be discussed there.

In the general case where there is no such parity, it is generally impossible to diagonalize simultaneously $y_1,~y_2$ and the Higgs mass matrix, and we then see that a Higgs LFV operator is already present in the renormalizable Lagrangian in the Higgs mass basis.
In order to connect this result with our effective theory framework, one must integrate out the other physical scalars (assumed heavier than the $125~$GeV Higgs). Since, by eq. (\ref{f:2HDMrot}) the physical scalar $h$ lives in both doublets, the contributions coming from $\Phi_2$ proportional to $y_2 \cos(\alpha-\beta)$ to the Yukawa matrix do not decouple for big $m_H$ and are not suppressed by the new physics scale, being in general stronger than the higher dimensional operators.

Recently, the authors of \cite{sierra} have shown that in such a case one may obtain the $h \rightarrow \tau \mu$ process at an observable level, despite the limits from radiative decays. The reason being that the Higgs LFV is generated by a renormalizable operator, at tree level, while the tau radiative decay is loop suppressed. Furthermore, in certain regions of parameter space, a cancellation between the heavy scalar and pseudoscalar becomes possible, leading to additional suppression of the radiative decay.


\subsection{SUSY}
\label{subsecSUSY}

In supersymmetric models such as the Minimal Supersymmetric Standard Model (MSSM), anomaly cancellation and holomorphy force one to include two Higgs doublets of opposite Hypercharge, that give mass separately to up type and down type fermions. This corresponds then to a type II 2HDM, at least at tree level.

\begin{figure}
\begin{subfigure}{.5\textwidth}
    \center
    \includegraphics[width=.6\linewidth]{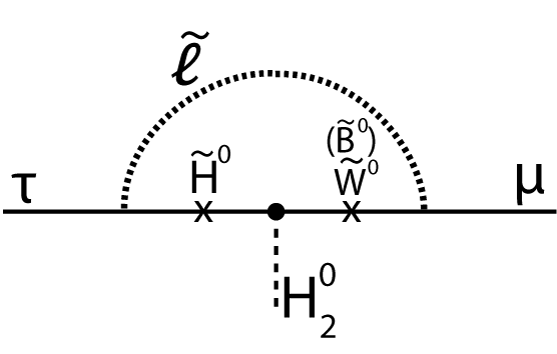}
    \caption{}
  \label{fig:susy1}
\end{subfigure}%
\begin{subfigure}{.5\textwidth}
\center
   \includegraphics[width=.6\linewidth]{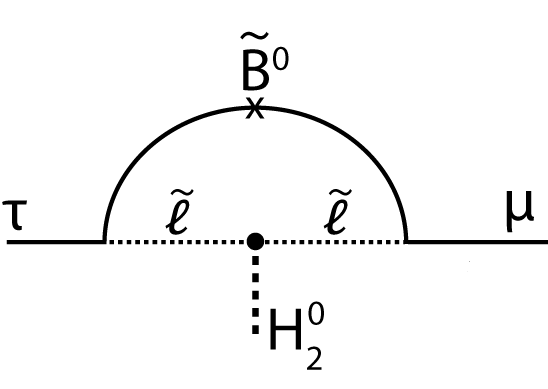}
   \caption{}
  \label{fig:susy2}
\end{subfigure}
\begin{subfigure}{1.\textwidth}
\center
   \includegraphics[width=.3\linewidth]{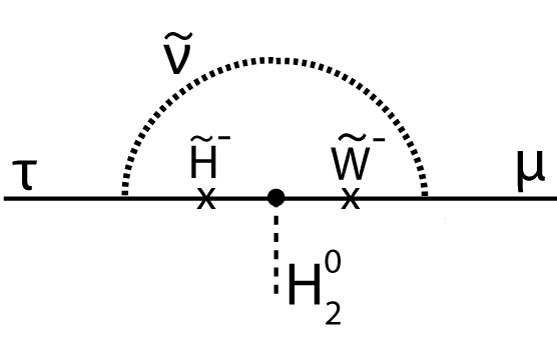}
   \caption{}
  \label{fig:susy3}
\end{subfigure}%
%
\caption{Diagrams leading to coupling between charged leptons and the ``wrong'' Higgs doublet (i.e., the one not responsible for their mass at tree level). The slepton, higgsino, wino and bino fields are denoted $\tilde{l}~,\tilde{H}^{0,\pm}_{1,2}~,\tilde{W}^{0,\pm}~\mbox{and}~\tilde{B}^0~$, respectively, and the cross denotes a mass insertion. Chirality labels are suppressed for simplictity.}
\label{susyfigs}
\end{figure}

At one-loop order, one may generate Yukawa couplings with the ``wrong'' Higgs boson, as in the diagrams of Fig. \ref{susyfigs}. Consider the soft SUSY breaking trilinear terms, and slepton masses
\begin{equation}
\mathcal{L}_{\mathrm{soft}} = -A_e \tilde{E} \tilde{L} H_d+\mathrm{h.c.}-M_L^2 |\tilde{L}|^2-M_E^2 |\tilde{E}|^2,
\end{equation}
where $\tilde{E},~ \tilde{L}$ are the left and right slepton fields and $H_d$ is the down type Higgs doublet. The trilinear couplings $A_e$  and the soft masses $M_L,~M_E$ are matrices in flavor space that in general are not aligned both with each other and the Yukawa matrices. This allows the diagrams of Fig. \ref{susyfigs} to generate small flavor violating transitions. Adding a photon to the loop, and exchanging the Higgs for its vev, one may then generate the radiative lepton decay with a Wilson coefficient that is directly proportional to the Higgs LFV one. Just as in the composite Higgs case, both Wilson coefficients, $c_{6H}$ and $c_{6\gamma}$ are correlated, however, unlike in that scenario, they are both generated at \emph{the same loop order}\footnote{ While the diagrams of Fig. \ref{susyfigs} are directly related to the LFV Higgs decays, there are other classes of diagrams that contribute to $\tau \rightarrow \mu \gamma$, as in \cite{susydiagramas}, and many more if R-Parity is violated, see \cite{Arhrib:2012ax}.}.

Since $\efOp{6\gamma}$  contributes to $\tau \rightarrow \mu \gamma$ at tree level, $c_{6\gamma}$ will strictly bounded. The relation between $c_{6H}$ and $c_{6\gamma}$ then implies $\mbox{BR}(h\rightarrow\tau\mu)$ to be very small. The authors of \cite{susydiagramas}, for instance, obtain $\mbox{BR}(h\rightarrow\tau\mu) \lesssim 10^{-4}$ from $\mbox{BR}(\tau\rightarrow\mu\gamma) < 3.1 \times 10^{-7}$ (which is an older constraint, superseded  by the one shown in Table \ref{tab:exp}). A R-Parity violating scenario was explored in \cite{Arhrib:2012ax} with $\mbox{BR}(h\rightarrow\tau\mu) \lesssim 10^{-5}$ as a constraint.

This puts the LFV Higgs decays at least two orders of magnitude below the current experimental sensitivity, and disfavors this class of models in case the current excess turns out to be signal.


\subsection{Higgs Portal}
\label{subsecPortal}

One class of models where radiative lepton decay may be 
small enough to satisfy bounds and still allow observable
Higgs LFV decays is given by the Higgs portal models, where a scalar singlet odd under a $\mathsf{Z}_2$ symmetry is added to the SM.
The coupling terms of this field to the Higgs and leptons, as well as its self couplings, are given by the effective Lagrangian:

\begin{equation}
\mathcal{L}_{SH}= \frac{\xi}{2} S^2 H^\dag H -\frac{\tilde{y}}{2\Lambda^2}S^2 \bar{L}H E-\frac{m_{S_0}^2}{2} S^2 +\frac{\lambda_S}{4!} S^4,
\label{higpot}
\end{equation}
where $S$ is the singlet field, $\xi$, $\tilde{y}$ are coupling constants ($\tilde{y}$ is a matrix in flavor space), and $\Lambda$ is a scale associated with some heavier sector (e.g., vector-like leptons or an additional scalar weak doublet) which generates the effective interaction of $S$ with SM fermions, not to be confused with the mass of $S$, $m_S$. The flavor violating decays are generated by the $\tilde{y}$ couplings, which may be misaligned with respect to the SM Yukawas.

The mass of $S$ is given at tree level by
\begin{equation}
m_S^2=m_{S_0}^2+ \frac{\xi}{2} v^2.
\end{equation}
At one loop order, there are additional contributions from $\lambda_S$ and $\tilde{y}$ which we neglect.

\begin{figure}
\begin{subfigure}{.5\textwidth}
    \center
    \includegraphics[width=.5\linewidth]{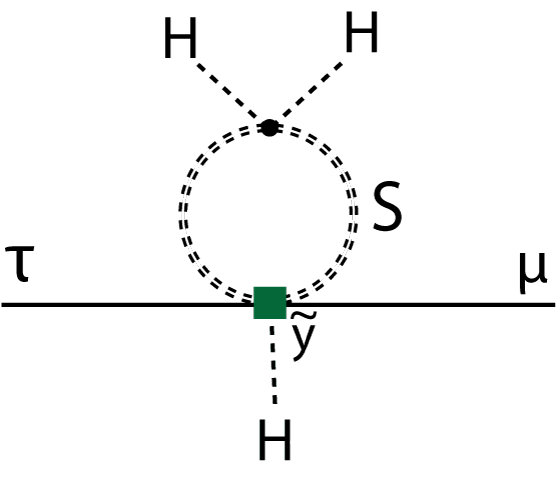}
    \caption{}
  \label{fig:portala}
\end{subfigure}%
\begin{subfigure}{.5\textwidth}
    \center
   \includegraphics[width=.5\linewidth]{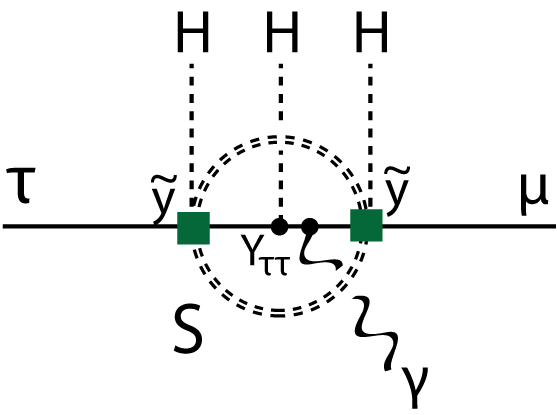}
   \caption{}
  \label{fig:portalb}
\end{subfigure}
\caption{
    LFV diagrams for a Higgs Portal model,  (a) Higgs LFV interaction at one loop. (b) Radiative lepton flavor violation at two loops.}
\label{portalfigs}
\end{figure}

In this class of models, the Higgs LFV interaction is generated at one loop order by the diagram of Fig. \ref{fig:portala}, which becomes $\efOp{6H}$ if we also integrate out $S$.

The radiative lepton flavor violation is generated only at \emph{two-loop} order by the sunset diagram of Fig. \ref{fig:portalb}. Both $\efOp{6 \gamma}$ and $\efOp{8 \gamma}$ are obtained in the large $m_S$ limit,  $\efOp{8 \gamma}$ is directly obtained from Fig. \ref{fig:portalb}, while $\efOp{6 \gamma}$ is obtained by closing a third loop with two of the three Higgs lines in that diagram, making this operator even more suppressed.

We note that, while from a low energy point of view, $m_S$ is the cutoff scale that determines the suppression of the irrelevant operators, an explicit calculation from the diagrams of Figs. \ref{fig:portala} and \ref{fig:portalb} shows that $c_{6H}$  and $c_{8\gamma}$ depend on $m_S$ only logarithmically, becoming important for a larger gap between $m_S$ and the scale $\Lambda$, see below\footnote{For a large gap, of order $m_S/\Lambda \lesssim 0.37$, perturbativity is lost and one must resum the large logarithms with the RG. This is beyond the scope of our present analysis.}.

Because the radiative decays are generated at two loops, they are naturally suppressed with respect to the Higgs LFV rates. Furthermore, since $S$ is uncharged, it is impossible to add a photon to the loop of Fig. \ref{fig:portala}, and the Wilson coefficients $c_{6H}$ and $c_{8\gamma}$ are uncorrelated. In this case, we find that the conclusions of \cite{harnik} apply.

We estimate the size of the contributions to $h\rightarrow \tau \mu$ and $\tau \rightarrow \mu \gamma$ to be, respectively:
\begin{align}\label{f:portal}
&c_{6H}\frac{v^2}{\Lambda^2} = \frac{\xi \tilde{y}_{\tau \mu}}{32 \pi^2}\frac{v^2}{\Lambda^2}\left( \log{\frac{m_S^2}{\Lambda^2}}+1-2 \sqrt{\frac{2 m_S^2}{m_h^2}-1}\;\sin^{-1}\left(\frac{m_h}{2 m_S}\right)\right), \nonumber \\
&c_{8\gamma}\frac{e v^3}{\Lambda^4} \simeq \sum_{k=\tau,\mu}e\frac{\tilde{y}_{\tau \mu}\tilde{y}_{kk}  m_\tau}{2(16 \pi^2)^2}\frac{v^2}{\Lambda^4}\log^2{\frac{m_S^2}{\Lambda^2}},
\end{align}
where the result for $c_{8\gamma}$ is only a leading order estimate.

\begin{figure}
\begin{subfigure}{.5\textwidth}
    \center
    \includegraphics[width=.9\linewidth]{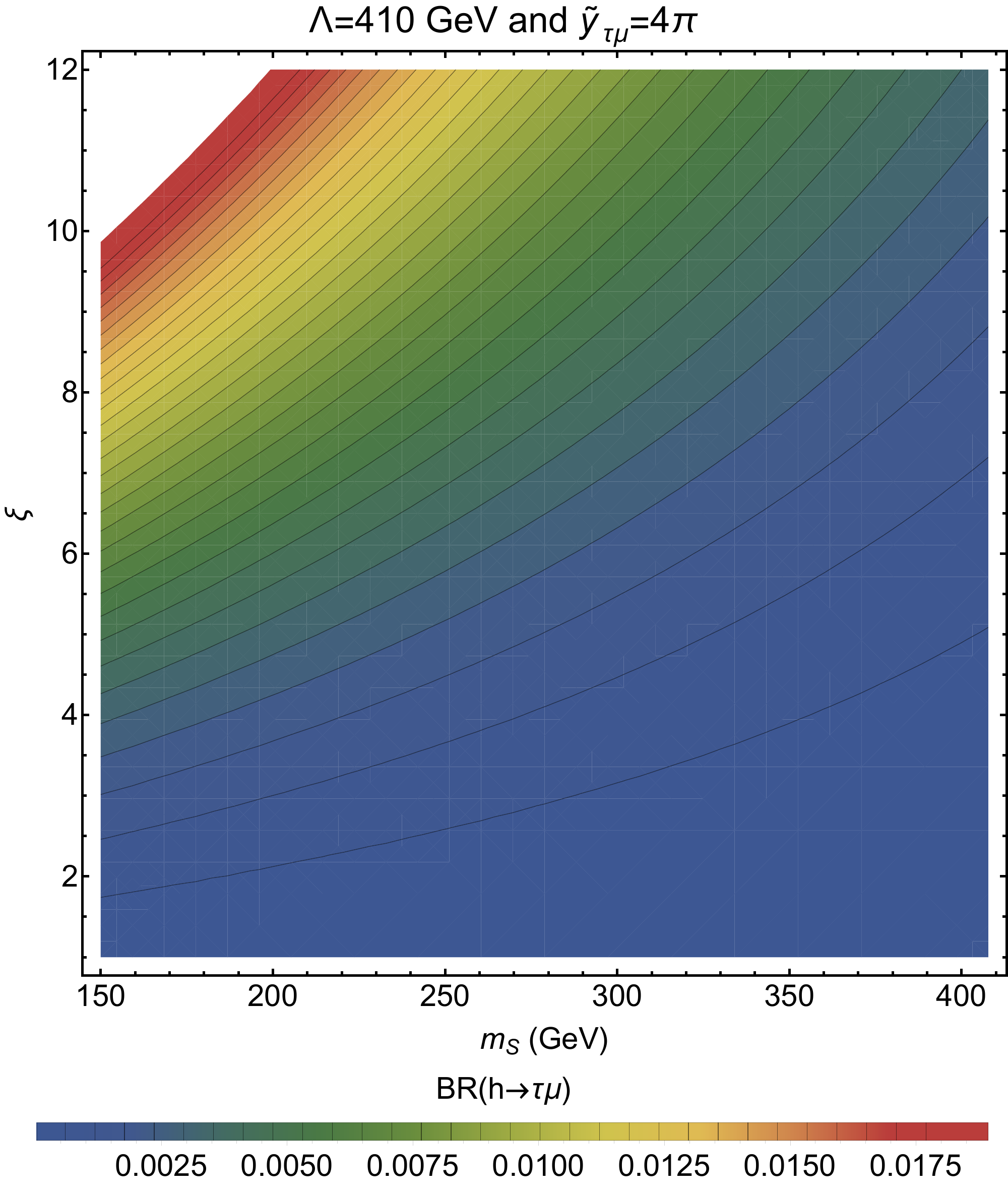}
    \caption{}
  \label{fig:BR}
\end{subfigure}%
\begin{subfigure}{.5\textwidth}
    \center
  \includegraphics[width=.9\linewidth]{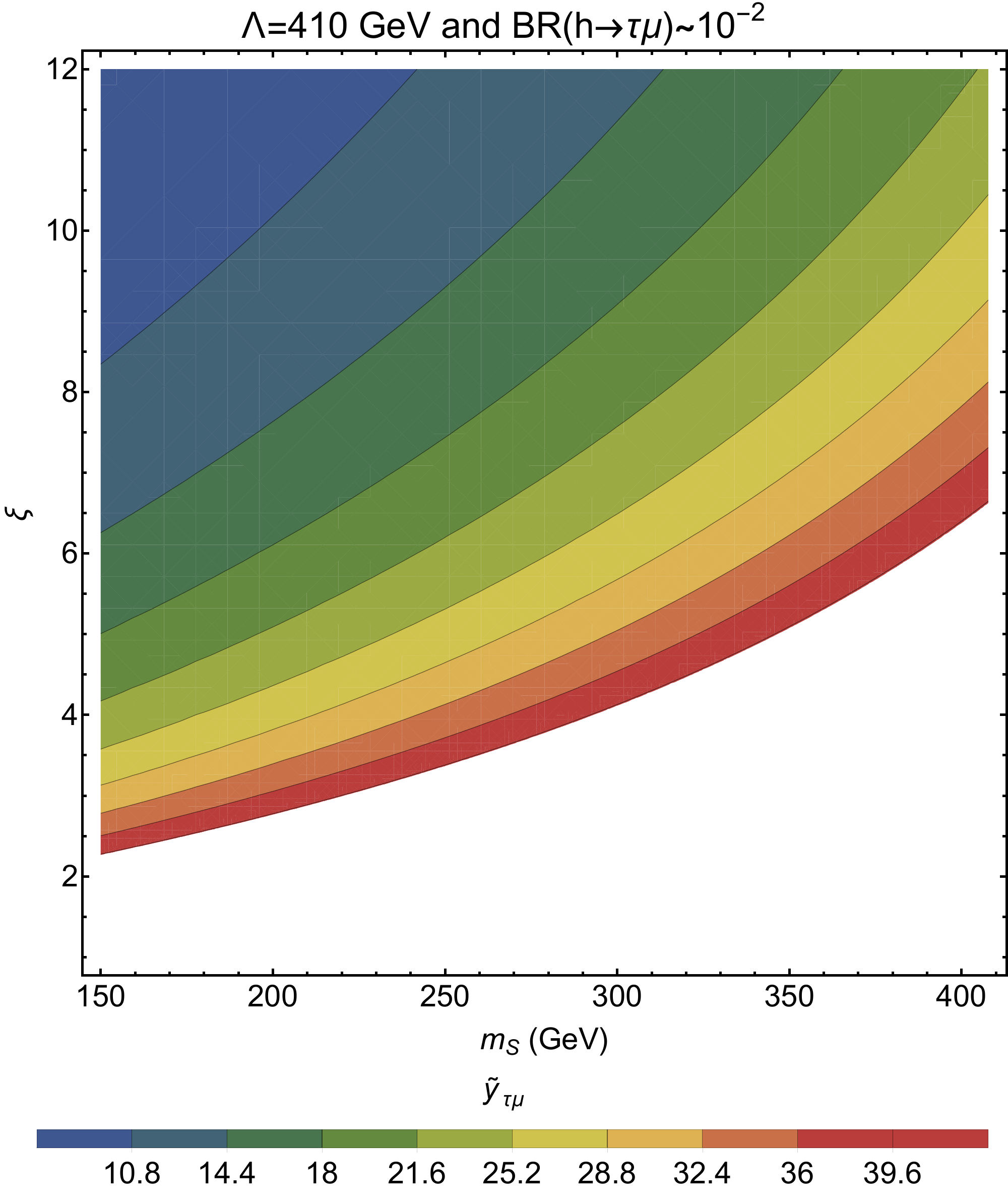}
   \caption{}
  \label{fig:y}
\end{subfigure}
\caption{Higgs portal coupling $\xi$ as a function of the scalar mass $m_S$. We show contour lines for (a) BR$(h\rightarrow \tau \mu)$ with $\tilde{y}_{\tau \mu}=4\pi$ and for (b) $\tilde{y}_{\tau \mu}$ with BR$(h\rightarrow \tau \mu)=10^{-2}$.
The region $\Lambda= 410~$GeV  and $\tilde{y}_{\tau\mu}\leq 4\pi$ satisfies perturbativity bounds (see text).}
\label{grafportal2}
\end{figure}

 Using eqs. (\ref{f:portal}), we can put bounds on the parameter space of this simple model. The constraints are that the radiative $\tau$ decay be below the experimental rate of Table \ref{tab:exp}, that the gap between $m_S$ and $\Lambda$, as well as the size of the couplings $\xi$ and $\tilde{y}_{\tau \mu}$ be small enough to mantain perturbative control. 
 Explicitly, we use the perturbativity contraints
\begin{align}
&\xi \tilde{y}_{\tau \mu} \lesssim 32 \pi^2 \frac{\Lambda^2}{v^2} \simeq 315.8 \frac{\Lambda^2}{v^2},\nonumber \\
&1>\frac{m_S}{\Lambda} \gtrsim E^{-1} \simeq 0.37.
\end{align}
In the second equation, $E=2.718\ldots$ is the Euler number. We note that no stringent bound on $\tilde{y}_{\tau \mu}$ is implied by $c_{8\gamma}$, because of the $\tau$ Yukawa suppression.
Imposing these constraints, we obtain the plots of Fig. \ref{grafportal2}.

One sees that, satisfying the constraints, it is possible to generate a Higgs LFV decay rate of about $BR(h\rightarrow \tau \mu)=10^{-2}$, observable at the LHC, while the radiative tau decay is well under control, with a rate of order $BR(\tau\rightarrow \mu \gamma)<1.5\times 10^{-17}$, where the maximum value is obtained for couplings saturating the perturbativity bound. However, if $S$ was stable (being the lightest $\mathsf{Z}_2$ odd particle), it's thermal relic abundance would have to be below or equal to the observed Dark Matter abundance, $\Omega_S\leq \Omega_{DM}=0.227$, implying $\xi \leq 3 \times 10^{-4} (m_S/1~\mbox{GeV})$ \cite{dmabundance1,dmabundance2}, which is
far too small to allow $c_{6H}$ to be sizeable.
One way to avoid this is to extend the scalar sector to obey a larger global symmetry than $\mathsf{Z}_2$. In this case, only the lightest scalar must satisfy the DM constraints, while it's heavier partners may have larger couplings to the Higgs, of order $\xi \simeq 10$ and so allow the $h \rightarrow \tau \mu$ rate to be generated at an observable level.

This class of models then allows one to evade the bound on $\tau \rightarrow \mu \gamma$ and still generate $h \rightarrow \tau \mu$ at a rate near the current experimental sensitivity. A signal of this decay would then be indicative that this kind of model may be playing a role.

Detailed phenomenology of the scalar sector and the  viability of the lightest scalar as a DM candidate will be presented elsewhere, however, see \cite{lee}.


\section{Conclusions}

We have examined the possibility of a new physics contribution to the Higgs LFV process $h \rightarrow \tau \mu$, which has been recently constrained by CMS to $\mbox{BR}(h\rightarrow\tau\mu) < 1.57\%$ at $95\%$ or that may be rather seen as a signal, with $\mbox{BR}(h\rightarrow\tau\mu) = \left( 0.89 \substack{+0.40 \\ -0.37}\right)\%$ \cite{cms}.

From a model independent perspective, in order to obtain a signal big enough to be detected or restricted by the LHC ($\mbox{BR}(h\rightarrow \tau \mu) \sim 10^{-2}$) one must consider
models in which the Higgs LFV operator $\hat{O}_{6H}$  is sizeable but does not violate the important constraint from
radiative lepton decay, $\mbox{BR}(\tau \rightarrow \mu \gamma)< 4.4 \times 10^{-8} $ \cite{babar}. This usually requires the
contribution of the  dipole operators $\hat{O}_{6\gamma},~\hat{O}_{8\gamma}$ to the tau decay rate to be negligible, which implies that they are generated at a higher loop order than $\hat{O}_{6H}$ by the UV continuation of the SM. The dipole operators, specially $\hat{O}_{8\gamma}$, have been generally neglected in model independent approaches to LFV Higgs decays. We show that the contributions of these operators can be important and thus put strong bounds on $\hat{O}_{6H}$.

Looking at specific classes of models found in the literature, we see that typically the Wilson coefficients of $\hat{O}_{6\gamma},~\hat{O}_{8\gamma}$ are correlated with $\hat{O}_{6H}$, and make it difficult to avoid the radiative decay bound and still get Higgs LFV at an observable level. This is the case of SUSY, where $\hat{O}_{6H}$ is generated at one-loop order, and Composite Higgs models, in which it is obtained at tree level. In both these cases, one may add a photon to $\hat{O}_{6H}$ and thus get $\hat{O}_{6\gamma}$ with a similarly sized Wilson coefficient. For this reason, these kinds of models are disfavored by the data, should the measured rate for $h \rightarrow \tau \mu$ be confirmed as a signal.

On the other hand, there are certain kinds of models in which one may avoid this conclusion. Examples are a type III Two Higgs Doublet Model, where Higgs LFV is produced by a renormalizable operator,
and models with an extended, gauge singlet scalar sector, in which case the tau radiative decay only appears via $\hat{O}_{8\gamma}$, generated at two loops, and is negligible. These options for extending the Standard Model may be favored, should this signal persist.

More generally, any model aiming to explain the data should have a mechanism to suppress $\tau \rightarrow \mu \gamma$ independently from Higgs LFV, by guaranteeing that the dipole operators are generated at a higher loop order than $\hat{O}_{6H}$.

\begin{acknowledgments}
This work has been supported by Funda\c c\~ao de Amparo \`a Pesquisa do Estado de S\~ao Paulo (FAPESP) under contracts 2012/01890-7,  2012/21436-9, 2012/21627-9 and 2013/03862-3.
The authors would like to thank Eduardo Pont\'{o}n and Andr\'{e} Lessa for useful discussions.
\end{acknowledgments}


\end{document}